\documentclass[conference]{IEEEtran}
\IEEEoverridecommandlockouts
\usepackage{cite}
\usepackage{amsmath,amssymb,amsfonts}
\usepackage{algorithmic}
\usepackage{graphicx}
\usepackage{subcaption}
\usepackage{textcomp}
\usepackage{float}
\usepackage{comment}
\renewcommand{\figurename}{Figure}
\usepackage[hidelinks]{hyperref}
\usepackage{xcolor}

\usepackage{caption}

\usepackage{dblfloatfix}

\def\BibTeX{{\rm B\kern-.05em{\sc i\kern-.025em b}\kern-.08em
    T\kern-.1667em\lower.7ex\hbox{E}\kern-.125emX}}
\begin{document}

%\title{Conference Paper Title*\\

%Quantum Circuit Assembler: A Strategy for Improving Job Executions on Quantum Machines

%Optimizing Job Execution on Quantum Machines with Circuit Assembly

\title{Quantum circuit scheduler for QPUs usage optimization}

\makeatletter % changes the catcode of @ to 11
\newcommand{\linebreakand}{%
  \end{@IEEEauthorhalign}
  \hfill\mbox{}\par
  \mbox{}\hfill\begin{@IEEEauthorhalign}
}
\makeatother % changes the catcode of @ back to 12

\author{\IEEEauthorblockN{Javier Romero-Alvarez}
\IEEEauthorblockA{\textit{Quercus Software Engineering Group}\\
\textit{Universidad de Extremadura}\\
Cáceres, Spain \\
0000-0002-3162-1446}
\and
\IEEEauthorblockN{Jaime Alvarado-Valiente }
\IEEEauthorblockA{\textit{Quercus Software Engineering Group}\\
\textit{Universidad de Extremadura}\\
Cáceres, Spain \\
0000-0003-0140-7788}
\and
\IEEEauthorblockN{Jorge Casco-Seco}
\IEEEauthorblockA{\textit{Quercus Software Engineering Group}\\
\textit{Universidad de Extremadura}\\
Cáceres, Spain \\
0009-0003-9166-6827}
\and

\linebreakand

\IEEEauthorblockN{Enrique Moguel}
\IEEEauthorblockA{\textit{Quercus Software Engineering Group}\\
\textit{Universidad de Extremadura}\\
Cáceres, Spain \\
0000-0002-4096-1282}
\and
\IEEEauthorblockN{Jose Garcia-Alonso}
\IEEEauthorblockA{\textit{Quercus Software Engineering Group}\\
\textit{Universidad de Extremadura}\\
Cáceres, Spain \\
0000-0002-6819-0299}
\and
\IEEEauthorblockN{Javier Berrocal}
\IEEEauthorblockA{\textit{Quercus Software Engineering Group}\\
\textit{Universidad de Extremadura}\\
Cáceres, Spain \\
0000-0002-1007-2134}
\and

\linebreakand
\IEEEauthorblockN{Juan M. Murillo}
\IEEEauthorblockA{\textit{Quercus Software Engineering Group}\\
\textit{Universidad de Extremadura}\\
Cáceres, Spain \\
0000-0003-4961-4030}
}

\begin{comment}
\author{\IEEEauthorblockN{Anonymous author}
\IEEEauthorblockA{\textit{Anonymous affiliation}\\
\textit{Anonymous affiliation}\\
Anonymous address \\
orcid}
\and
\IEEEauthorblockN{Anonymous author}
\IEEEauthorblockA{\textit{Anonymous affiliation}\\
\textit{Anonymous affiliation}\\
Anonymous address \\
orcid}
\and
\IEEEauthorblockN{Anonymous author}
\IEEEauthorblockA{\textit{Anonymous affiliation}\\
\textit{Anonymous affiliation}\\
Anonymous address \\
orcid}
\and

\linebreakand

\IEEEauthorblockN{Anonymous author}
\IEEEauthorblockA{\textit{Anonymous affiliation}\\
\textit{Anonymous affiliation}\\
Anonymous address \\
orcid}
\and
\IEEEauthorblockN{Anonymous author}
\IEEEauthorblockA{\textit{Anonymous affiliation}\\
\textit{Anonymous affiliation}\\
Anonymous address \\
orcid}
\and
\IEEEauthorblockN{Anonymous author}
\IEEEauthorblockA{\textit{Anonymous affiliation}\\
\textit{Anonymous affiliation}\\
Anonymous address \\
orcid}
\and

\linebreakand
\IEEEauthorblockN{Anonymous author}
\IEEEauthorblockA{\textit{Anonymous affiliation}\\
\textit{Anonymous affiliation}\\
Anonymous address \\
orcid}
}

\end{comment}

\maketitle

\setlength{\parskip}{1mm}
\begin{abstract} %150-250

% Advances in the quantum domain are leading to numerous application possibilities in different sectors. 
Progress in the realm of quantum technologies is paving the way for a multitude of potential applications across different sectors. However, the reduced number of available quantum computers, their technical limitations and the high demand for their use are posing some problems for developers and researchers. Mainly, users trying to execute quantum circuits on these devices are usually facing long waiting times in the tasks queues. In this context, this work propose a technique to reduce waiting times and optimize quantum computers usage by scheduling circuits from different users into combined circuits that are executed at the same time. To validate this proposal, different widely known quantum algorithms have been selected and executed in combined circuits. The obtained results are then compared with the results of executing the same algorithms in an isolated way. This allowed us to measure the impact of the use of the scheduler. Among the obtained results, it has been possible to verify that the noise suffered by executing a combination of circuits through the proposed scheduler does not critically affect the outcomes.

\end{abstract}


%\textcolor{red}{Every “Regular Paper” manuscript can include 7 to 10 pages for the main contents (including all text, footnotes, figures, tables, and appendices) with additional pages for appropriate references.}

%%%%%%%%%%%%%%%%%%%%%%%%%%%%%%%%%%%%%%%%%%%%%%%%%%%%%%%%%%%%%%%%%%%%%%%%%%%%%%%%%%%%%%%
\section{Introduction}

\setlength{\parskip}{1.3mm}

In the emerging context of quantum computing, different perspectives on how to access quantum computers through cloud platforms are emerging. Leading providers such as Amazon, IBM, and Google have started entering the quantum computing field by investing large efforts and resources \cite{hassija2020present}. Service providers are offering access to quantum computers through the cloud, following a scheme similar to that used by traditional cloud computing. Through this new model, called Quantum Computing as a Service (QCaaS) \cite{islam2015review}, quantum software developers can use some of these technologies to run their code on real quantum hardware directly or by invoking them through services \cite{alvarado2023technological}.

%As well as offering developers the possibility of using these resources through their cloud platforms, in the form of services \cite{alvarado2023technological}. 

To improve these aspects, each year these companies manage to make progress in enhancing their quantum computers, allowing classical problems that take thousands of years to be solved in supercomputers or traditional computers to be solved in seconds using their quantum computers. In 2021, the first 127-qubit computer was announced by IBM, and one year later this number of qubits was increased considerably with the announcement of the first 433-qubit quantum computer. This quantum capacity is increasing significantly, with the latest announcement indicating that in 2025 IBM would complete the construction of a quantum computer with 4000 qubits \cite{choi2023ibm}.

However, even with this high capacity in terms of number of qubits, the average number of qubits used per circuit is currently 10'5 \cite{ichikawa2023comprehensive}. Therefore, a waste of resources has been observed in these providers, manifested in high costs, long waiting times, and limitations in the availability of quantum machines \cite{nguyen2024qfaas}. Mainly due to the fact that quantum computers are being underutilized in each of the executions by not using the maximum number of qubits they offer. This, mixed with the high demand for the use of quantum computers during specific time periods, means that developers wishing to use these devices will not be able to execute their tasks, usually in the form of quantum circuits, instantaneously or in a relatively reasonable amount of time.

These factors directly affect the efficiency of executions and the ability of users to obtain results promptly, where execution times can take from seconds to hours to complete. 

In contrast, in the field of classical High-Performance Computing, advanced parallelization and optimization techniques have been explored to improve performance and efficient resource utilization in traditional computational systems \cite{garg2002techniques}. These strategies have proven to be essential to maximize processing power and reduce execution times in computationally intensive tasks, allowing computers to work at the same time on different tasks, and offering users much shorter queuing times than we see in quantum computing.

In this context, and to address these challenges, this work propose to schedule circuits from different users, or from the same user, into combined circuits. This strategy aims to reduce waiting times, improve task execution, and reduce individual costs for developers. Efficiently joining circuits is also intended to increase the efficiency in the utilization of the resources available in quantum computers, making greater use of the qubits offered by the computers.

To validate this proposal, different widely known quantum algorithms have been selected. Their corresponding circuits has been executed on an IBM Quantum 127-qubit quantum processor, both individually as well as in different scheduled combinations. With this, it has been possible to verify that the proposed scheduler provides a substantial improvement for the developers in terms of reducing queue times and costs. Furthermore, the noise suffered by the execution of the combined circuits does not notably affect the results. 

To improve replicability, all the validation results, code, and data used in this work are available in a Zenodo public repository\footnote{Anonymous link}.

To present this, the rest of the paper is organized as follows: Section \ref{sec:Background} analyzes the background of the present work and discusses the most relevant related works. Section \ref{sec:Proposal} presents the proposal for unifying circuits using a quantum circuit scheduler. Section \ref{sec:Validation} demonstrates the feasibility of the proposal through the execution of different combinations of circuits. Finally, Section \ref{sec:conclusion} presents the conclusions and future works.

%%%%%%%%%%%%%%%%%%%%%%%%%%%%%%%%%%%%%%%%%%%%%%%%%%%%%%%%%%%%%%%%%%%%%%%%%%%%%%%%%%%%%%%%
\section{Background}\label{sec:Background}

%\textcolor{red}{- quantum software engineering \cite{moguel2022quantum}}

%\textcolor{red}{- Hybrid Quantum Computing and QCaaS}

%\textcolor{red}{- algun trabajo sobre la gestión de recursos en máquinas - iquantum \cite{Nguyen2023obm}}

%Stochastic Qubit Resource Allocation for Quantum Cloud Computing

%Which Quantum Circuit Mutants Shall Be Used? An Empirical Evaluation of Quantum Circuit Mutations

In the field of quantum computing, Quantum Software Engineering is beginning to be considered as a crucial field of research \cite{moguel2022quantum}. This is because it allows taking advantage of the capabilities of quantum systems to develop applications by making use of the knowledge of Classical Software Engineering. Thus, by integrating the principles of quantum mechanics into software design, it is possible to solve complex problems more effectively than with traditional approaches \cite{zhao2020quantum}.

Therefore, the concept of Hybrid Classical-Quantum systems has become relevant by combining elements of classical and quantum computation to improve the performance and scalability of these systems \cite{mccaskey2018hybrid}. So, this hybrid approach allows taking advantage of the strengths of both paradigms, which opens new opportunities to address problems in diverse fields.

In this context of advances in quantum computing, it is essential to address the challenges associated with resource management and queuing on quantum computers. As Quantum Software Engineering is starting to be used by different developers, it becomes necessary to develop solutions that optimize the utilization of available resources in quantum computing environments \cite{ichikawa2023comprehensive}. Specifically, the execution of quantum circuits in cloud service providers often involves significant waiting times before a task is completed, which can range from seconds to hours \cite{ravi2021quantum}. In addition, the current design of cloud providers does not allow for efficient use of quantum computer resources.
This is a recurring problem in quantum computing as a service \cite{garcia2021quantum} and will remain so as long as quantum computers remain exclusive to cloud providers, and it is expected to remain so for many years to come \cite{Nguyen2023obm}.

We are currently in the NISQ era \cite{leymann2020bitter}, where there are intrinsic problems in quantum computers when clustering circuits, such as noise, qubit decoherence \cite{brandt1999qubit}, interference \cite{murali2020software}, measurement problems \cite{nation2021scalable}, and even calibration failures \cite{wilson2020just}. In addition, since combining circuits to be jointly executed could experience more errors than the individual results due to an increased depth \cite{johnstun2021understanding}, it becomes essential to perform a technological validation of the proposal. This validation is addressed in the following sections of the article, to ensure that the combination of circuits does not significantly increase the noise in the results.

Several works are emerging to address this issue. For example, in Usandizaga, et al. \cite{usandizaga2023quantum} an empirical evaluation is proposed to help optimize the resources of quantum computers. To this end, they employ the quantum circuit mutation technique to generate defective benchmarks to evaluate the effectiveness of quantum software testing techniques.

On the other hand, Kaewpuang,  et al. \cite{kaewpuang2023stochastic} establish a two-stage stochastic programming model to minimize the total costs of quantum circuits under uncertainties in qubit requirements and quantum circuit waiting time. The presented system aims to enable users to acquire quantum computing resources through reservation and on-demand schemes and addresses the optimization of minimum quantum circuit waiting times.

In this aspect, the scheduling of circuits of different users in a single circuit, as proposed in this work, aims to address this issue by reducing waiting times, improving task execution, and reducing individual costs for developers.

\begin{figure*}[!ht]
    \centering
    \includegraphics[width=1\textwidth]{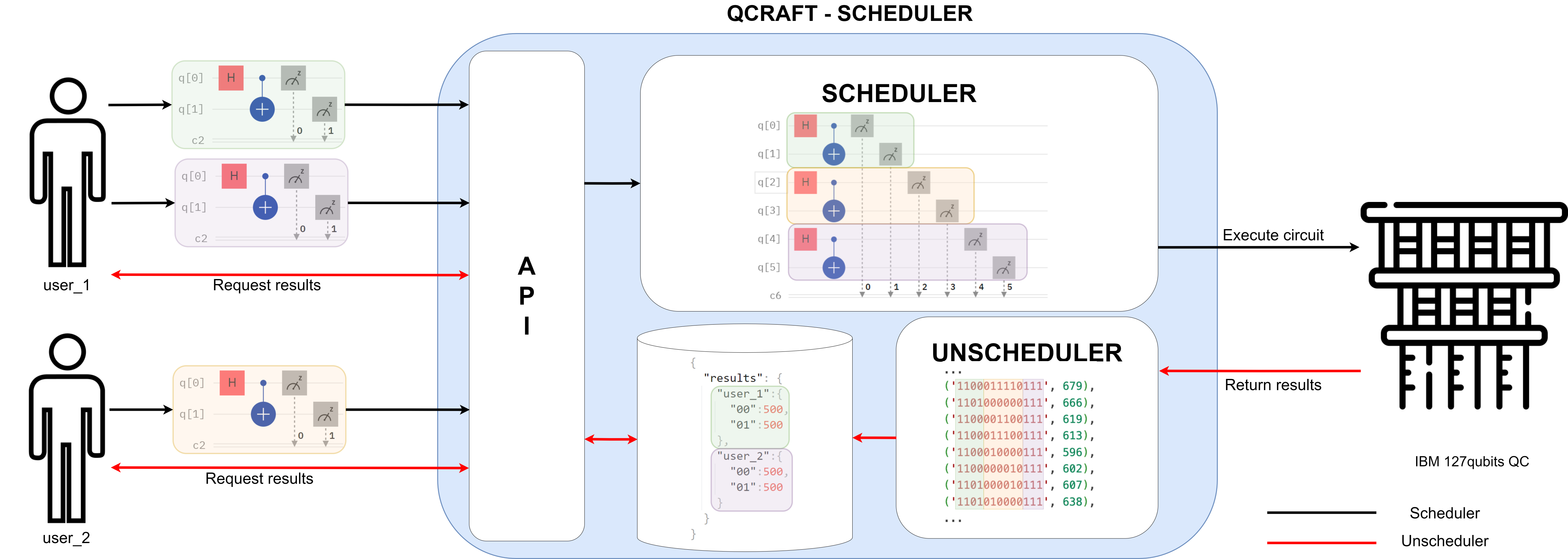}
    \caption{Process of the quantum circuit scheduler}
    \label{fig:composer}
\end{figure*}

The difference between these works concerning the proposal presented in this study is the approach to minimize waiting times and optimize the execution of tasks by scheduling circuits from different users in combined circuits. The main goal is to increase the efficiency in the utilization of the resources available in computers by scheduling circuits in an efficient way, which allows making greater use of the qubits offered by quantum machines. An additional aspect that distinguishes this proposal is its contribution to Hybrid Classical-Quantum systems, by reducing the waiting times it makes these architectures viable in more time constrained environments.

%\textcolor{red}{-necesidad de validar por el ruido}

%%%%%%%%%%%%%%%%%%%%%%%%%%%%%%%%%%%%%%%%%%%%%%%%%%%%%%%%%%%%%%%%%%%%%%%%%%%%%%%%%%%%%%%%%
\section{Unifying circuits with the Quantum Circuit Scheduler}\label{sec:Proposal}

This section provides a comprehensive explanation of the proposal for the scheduling of quantum circuits through a case study.

%\textcolor{red}{- explicar la propuesta de combinación de los circuitos} 

\subsection{Quantum Circuit Scheduler Proposal}\label{sectionComposer}

The full circuit scheduling process can be seen depicted in Figure \ref{fig:composer}. The proposed process starts with different users submitting their circuits to be executed to the scheduler API.

In their submissions, users specify the number of shots and submit the circuit via the URL generated by the Quirk circuit simulator\footnote{\url{https://algassert.com/quirk}} or via Python code, such as IBM's Qiskit language\footnote{\url{https://www.ibm.com/quantum/qiskit}}. Once received, the circuits are added to a queue, and each user receives a unique identifier for that execution.

Circuit executions are carried out in cycles. A cycle, in this context, can be defined as a finite period of time in which the scheduler will wait to receive more circuits to schedule until the number of available qubits is filled.

In this way, the system waits for one cycle to receive more circuits. If there are no more circuits in the queue, it proceeds with scheduling; otherwise, it restarts the cycle and continues waiting. During scheduling, the system loops through the elements in the queue and, as long as there are qubits available on the machine where the circuit is to be executed,  it combines them into a single circuit. If any circuit is too wide for this scheduling, it moves to the next in the queue.
Initially, it was considered for this proposal to allocate the a high number of shots per combined circuit, 10.000, which is imposed by the scheduler. 

Once scheduled, the combined circuit is run on the quantum machine. The results that this execution produces are a combination of all the results of each of the circuits that have been combined. Therefore, it is necessary to perform an unscheduling process of the results, to separate the results of each circuit individually. This process is carried out by the ``unscheduler''.

Once the execution and the unscheduling of the results have finished, users can check the results of their executions by making a new request to the API, sending as a parameter the unique identifier that was returned in the execution call.

%\textcolor{red}{- explicar por qué en ibm}

When carrying out the different executions of the circuits, IBM Quantum has been chosen as the provider. Specifically, their quantum processor ``Eagle'' has been utilized. This processor is developed and deployed to contain a total of 127 operational and connected qubits. Thus, being one of the processors currently available with a higher number of qubits, it allows for the scheduling of circuits that use a higher number of qubits and facilitates the analysis of the results obtained from the scheduled circuit executions, compared to the results obtained from their individual executions.

%\textcolor{red}{- hablar de los costes, tiempo...}

One of the main challenges in this kind of scheduling lies in the noise generated by quantum computers during calculations, which is given by the decoherence times of the qubits of each processor. The presence of disturbances can lead to the loss of qubit superposition, as well as changes in their state or phase \cite{chirolli2008decoherence}. Therefore, the longer the execution time and the complexity of the circuit, the greater the risk of these interferences occurring \cite{johnstun2021understanding}.

Therefore, this scheduler architecture and the proposed functionality need a technological validation and verification of how noise, decoherence, and other intrinsic factors of quantum computing affect circuit scheduling.

\subsection{Scheduling-Unscheduling Case Study}

Following the above proposal, for a better explanation, we detail the scheduling and unscheduling operation using a practical case study.

\begin{figure}[H]
    \centering
    \includegraphics[width=0.48\textwidth]{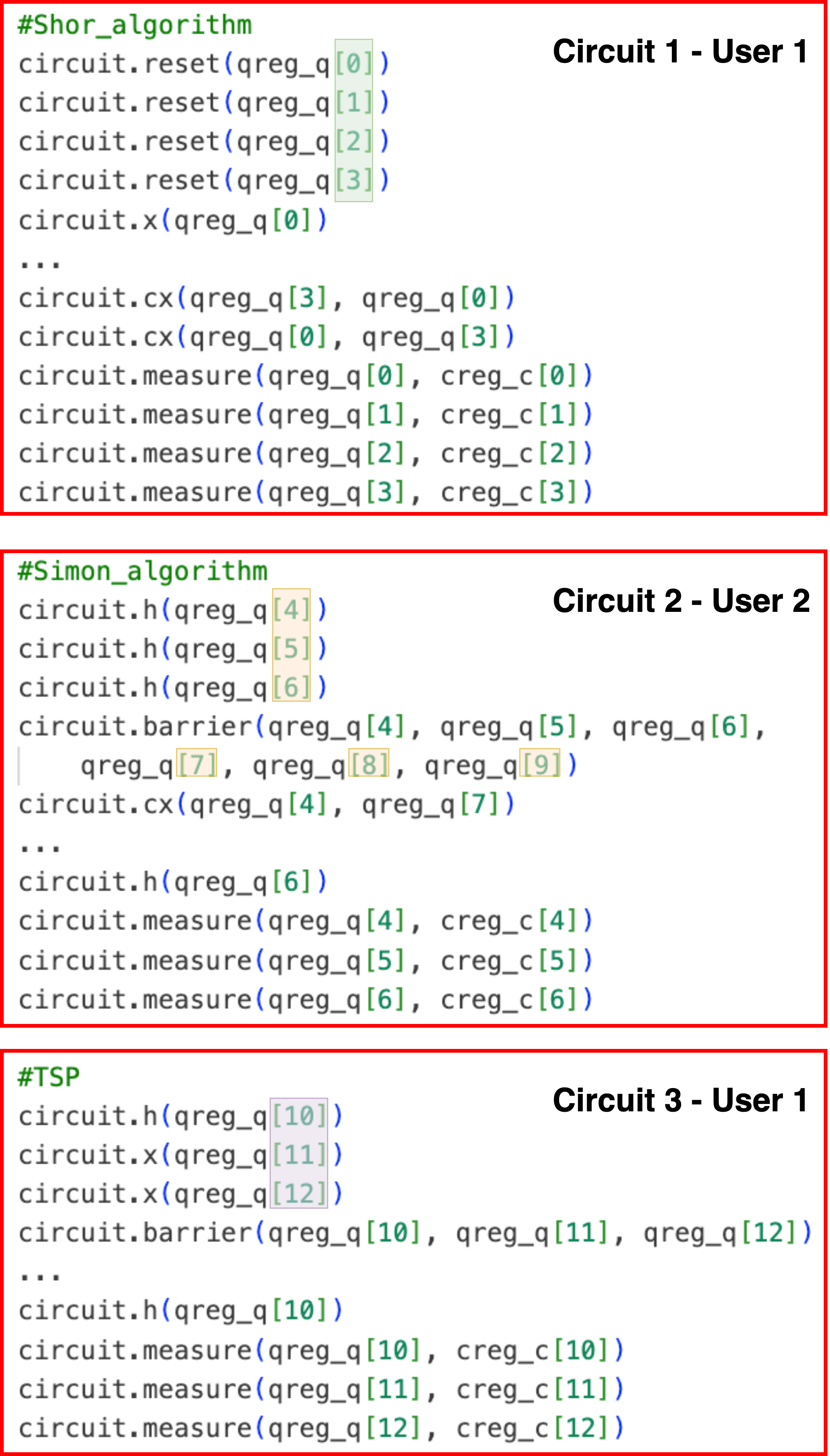}
    \caption{Quantum circuit scheduling case study}
    \label{fig:composedcircuit}
\end{figure}

This example consists of 2 users who will execute 3 circuits through the QCRAFT Scheduler (2 circuits for user 1, and 1 circuit for user 2). To facilitate the understanding of our proposal, we have selected the widely known Shor, Simon and TSP circuits, which, due to the number of qubits they use, allow us to visualise the qubit displacement that takes place in the scheduling. The code of these circuits is available at previously mentioned Zenodo repository.

%Vamos a utilizar como ejemplo el caso de tener 2 usuarios que envían a ejecutar 3 circuitos diferentes a través del QF4SE - Assembler. 

Following the process discussed in Section \ref{sectionComposer}, once users have submitted their circuits to the scheduler, these circuits encode the number of qubits they use from 0 to N, where N is the number of qubits used in total. 

These circuits go through the scheduling process, where they are modified in order to obtain a single circuit resulting from combining them into a single circuit. As can be seen in Figure \ref{fig:composedcircuit}, for a better understanding, it is possible to see a section of the circuits where the most relevant gates and the transpilation of the qubits in the circuit schedule can be observed. This transpilation implies the redefinition of the qubits to be used, i.e. if two circuits using 4 qubits each are joined, the first circuit will use the qubits between 0 and 3, and the second circuit will use the qubits between 4 and 7.

In Figure \ref{fig:composedcircuit} the numbers of the qubits that each circuit will use after going through the scheduling process are highlighted in different colours. 

\begin{itemize}
    \item User 1 sends Circuit 1 - Shor Algorithm: 4 qubits between 0 and 3.
    \item User 2 sends Circuit 2 - Simon Algorithm: 6 qubits between 4 and 9.
    \item User 3 sends Circuit 3 - TSP Algorithm: 3 qubits between 10 and 12.
\end{itemize}

The reference to the qubits used by each circuit is stored for later use in the unscheduling process. As can be seen in Figure \ref{fig:composedcircuitresults}, the result is a string of 13 bits, where:

\begin{itemize}
    \item The first 4 bits correspond to the results of Circuit 1 - User 1.
    \item The next 6 bits correspond to the results of Circuit 2 - User 2.
    \item The last 3 bits correspond to the results of Circuit 3 - User 1.
\end{itemize}

\begin{figure}[H]
    \centering
    \includegraphics[width=0.35\textwidth]{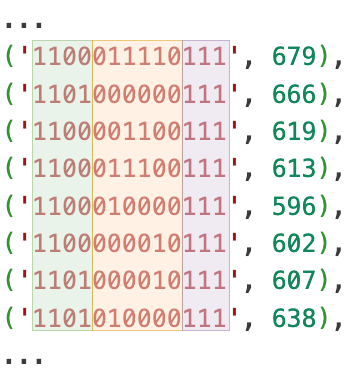}
    \caption{Partial example of quantum circuit unscheduling}
    \label{fig:composedcircuitresults}
\end{figure}

\begin{figure*}[!hb]
    \centering
    \begin{subfigure}[b]{0.45\textwidth}
        \centering
        \includegraphics[width=\textwidth]{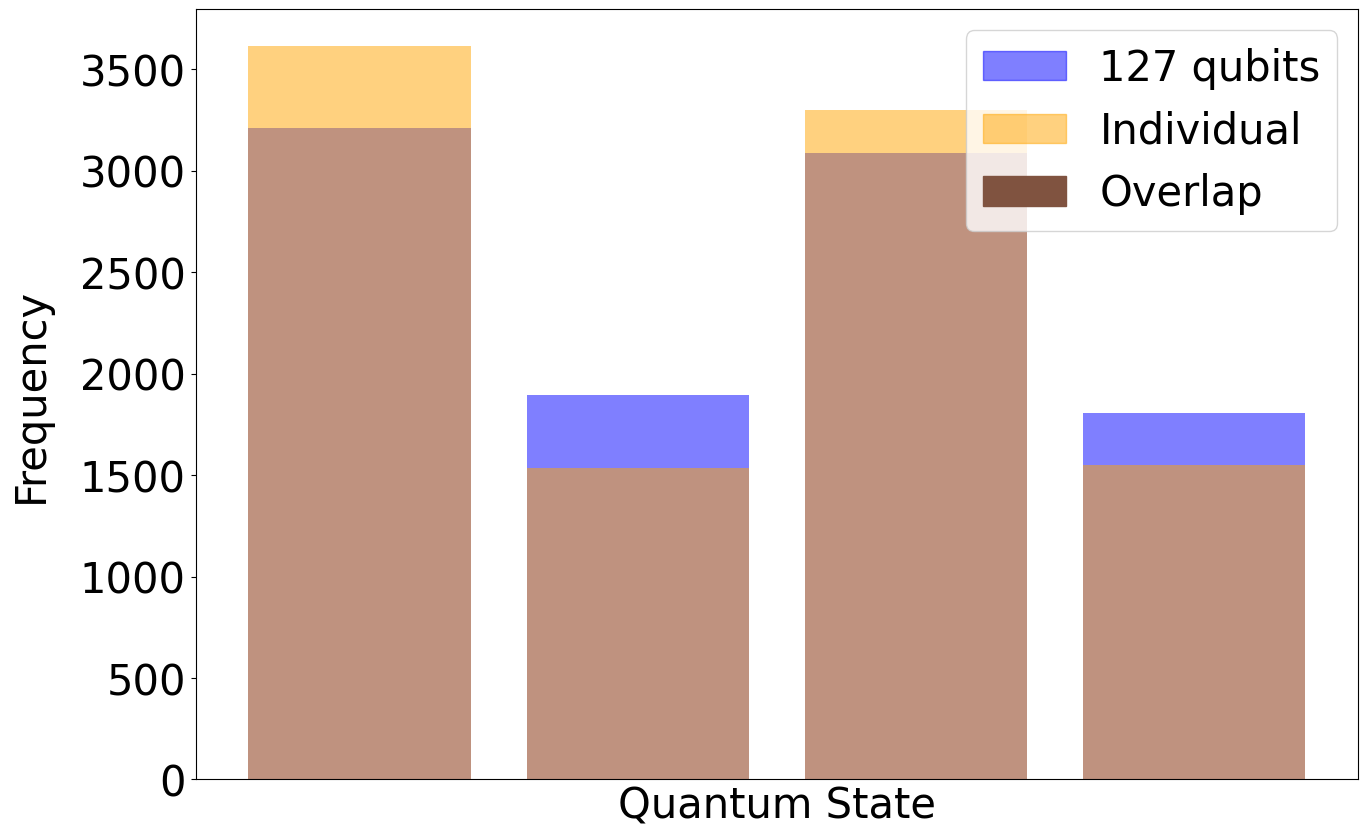}
        \caption{Results of QAOA algorithm}
        \label{fig:qaoa1}
    \end{subfigure}
    \hfill
    \begin{subfigure}[b]{0.45\textwidth}
        \centering
        \includegraphics[width=\textwidth]{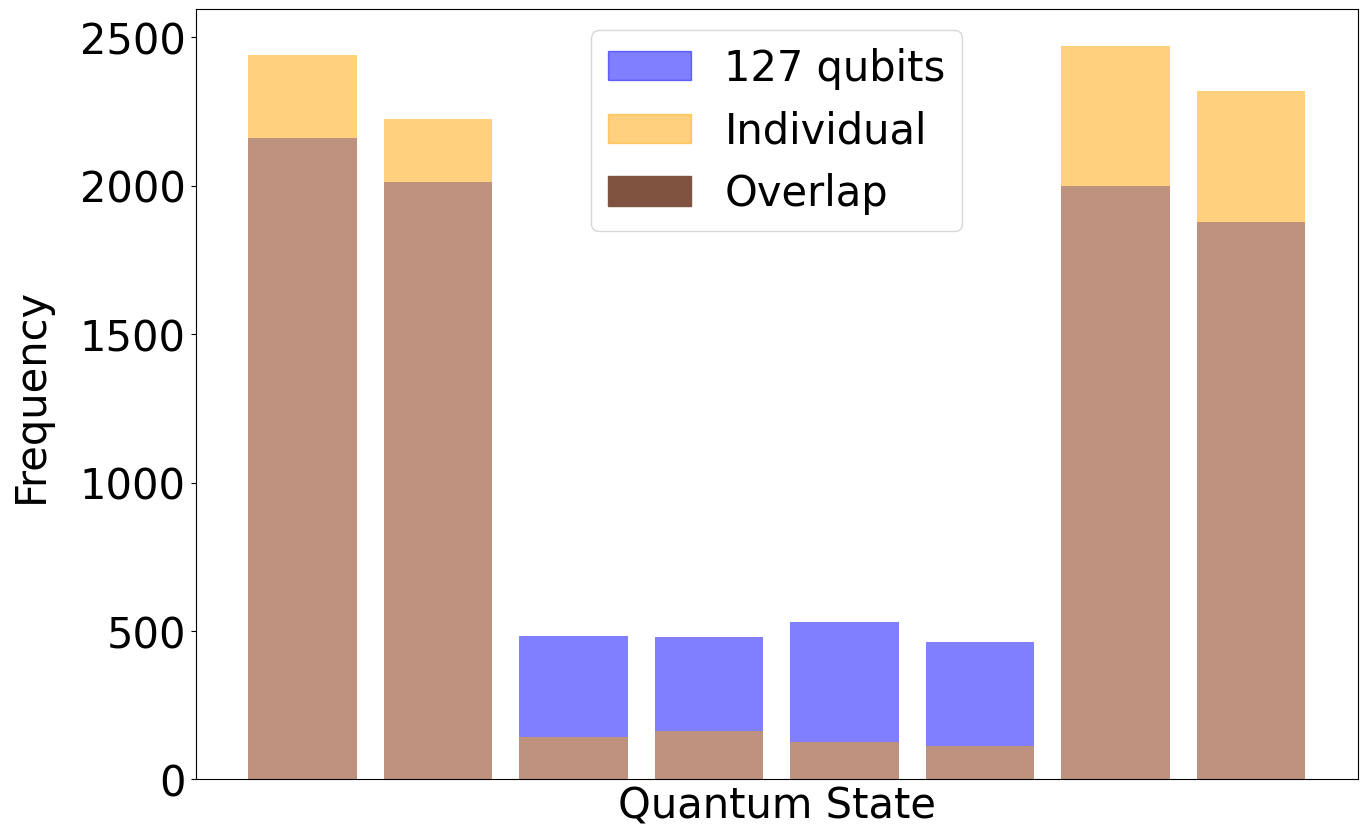}
        \caption{Results Simon algorithm}
        \label{fig:simon1}
    \end{subfigure}
    \vskip\baselineskip
    \begin{subfigure}[b]{0.45\textwidth}
        \centering
        \includegraphics[width=\textwidth]{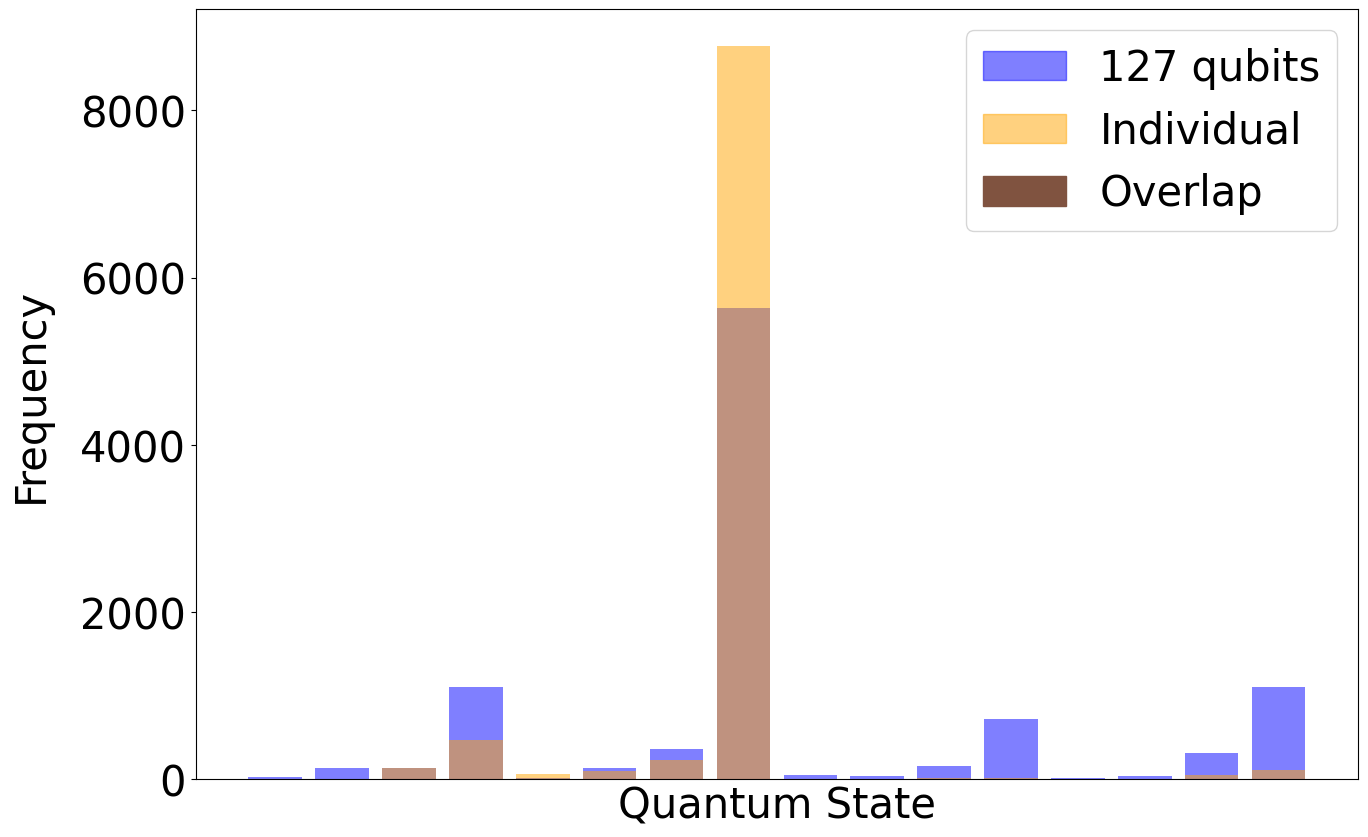}
        \caption{Results of Shor algorithm}
        \label{fig:shor1}
    \end{subfigure}
    \hfill
    \begin{subfigure}[b]{0.45\textwidth}
        \centering
        \includegraphics[width=\textwidth]{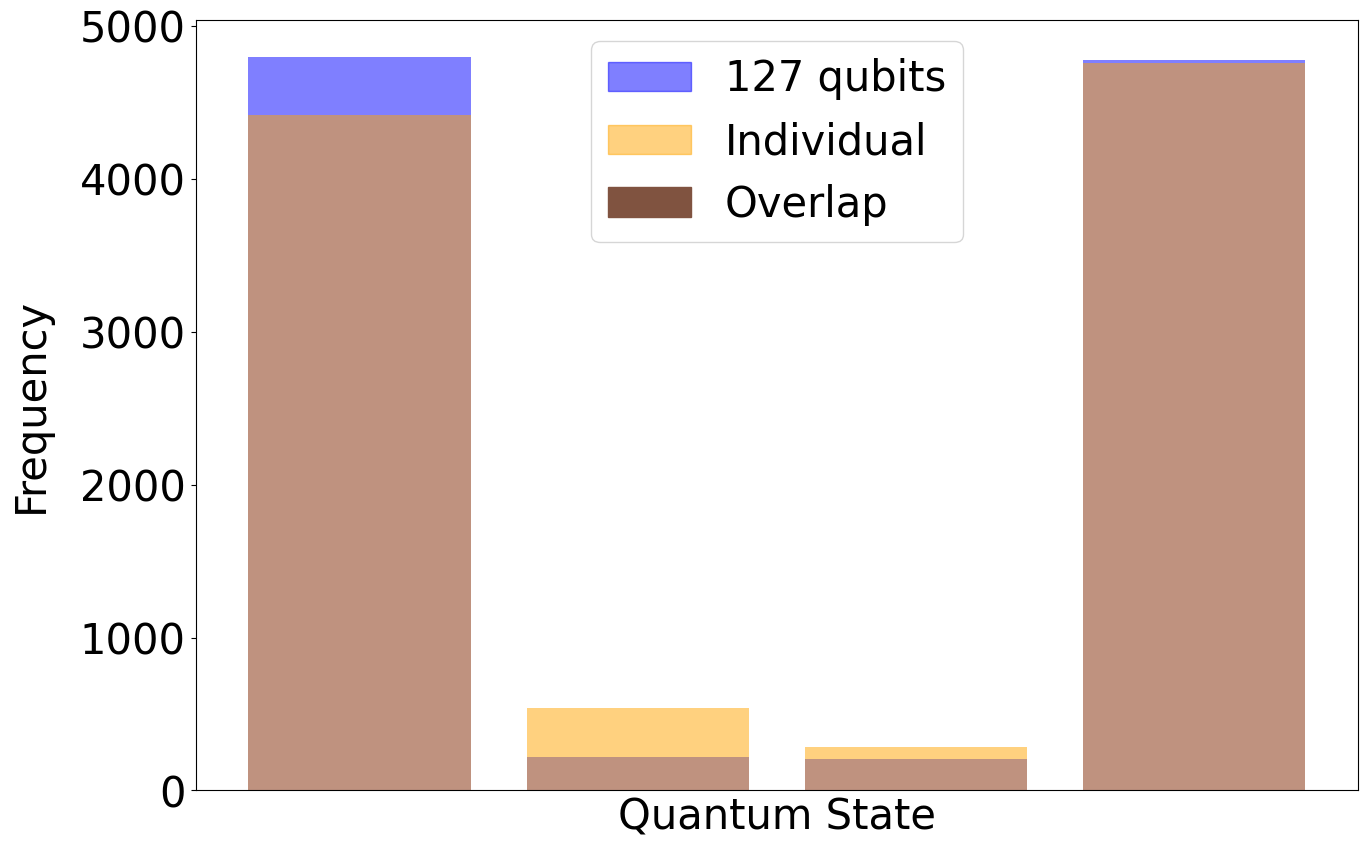}
        \caption{Results of Grover algorithm}
        \label{fig:grover1}
    \end{subfigure}
    \hfill%\vspace{3mm}
    \begin{subfigure}[b]{0.45\textwidth}
        \centering
        \includegraphics[width=\textwidth]{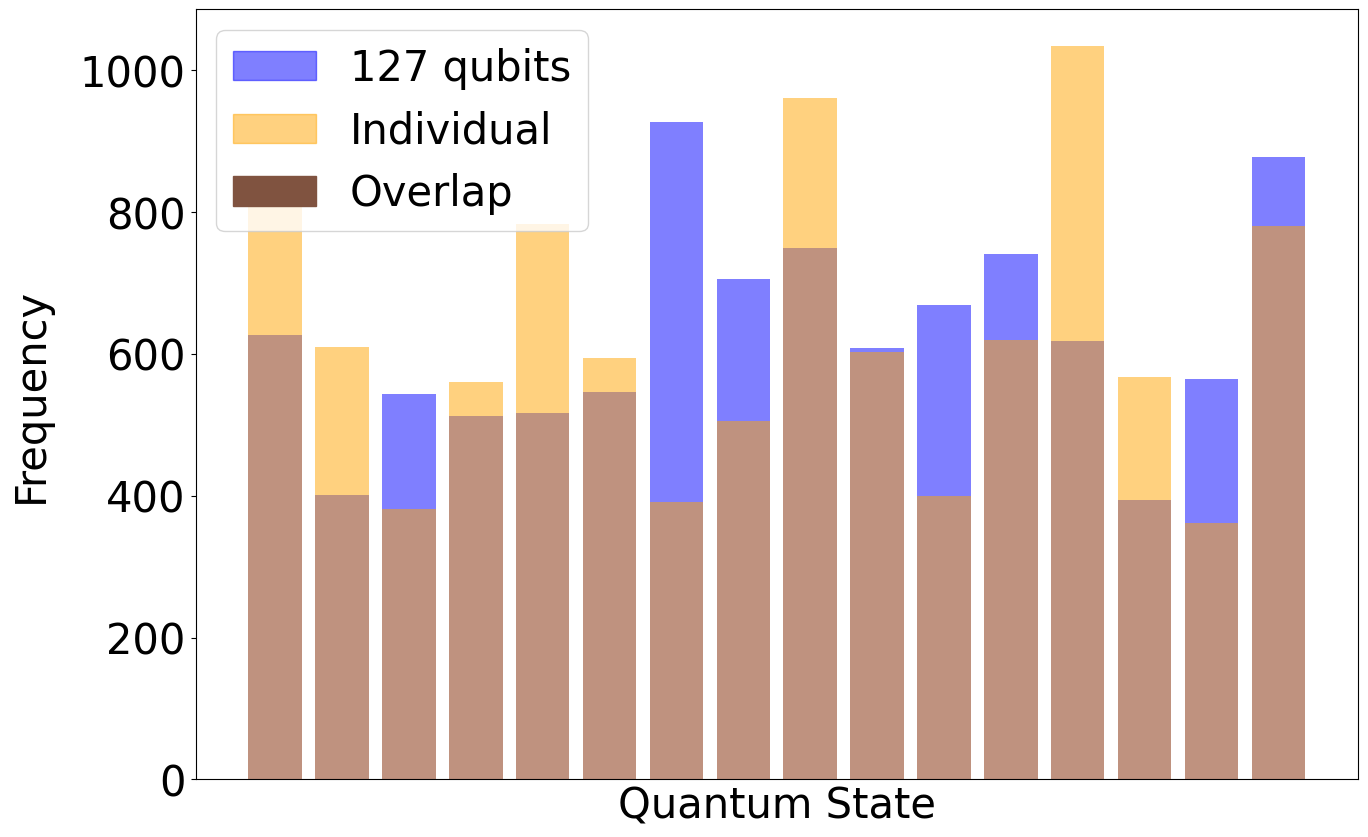}
        \caption{Results of Full Adder algorithm}
        \label{fig:fullAdder}
    \end{subfigure}
    \caption{Results of the execution of the circuits in schedule compared to individual}
    \label{fig:quantum_algorithms}
\end{figure*}

These results, once unscheduled, are stored in the scheduler, associated with the user's unique identifiers, and can be retrieved by the user through an API request.

%%%%%%%%%%%%%%%%%%%%%%%%%%%%%%%%%%%%%%%%%%%%%%%%%%%%%%%%%%%%%%%%%%%%%%%%%%%%%%%%%%%%%%%%%%%
\section{Validation and Results}\label{sec:Validation}

Technological validation is not only relevant, but also it is crucial to validate the proposal against noise and decoherence, thus recognizing the need to address these challenges. An environment without noise would ideally be optimal for quantum computing. However, in the NISQ era, errors are inevitable due to the current state of quantum computing, which is not yet sufficiently advanced.

\subsection{Scheduling of Quantum Algorithms}

%\textcolor{red}{- uso de algoritmos}

In our study, we have used a set of quantum circuits provided in the Zenodo public repository\footnote{Zenodo link}. This repository contains different implementations of circuits associated with important algorithms in the field of quantum computing. This variety has facilitated the realization of the process with circuits of different depths. To carry out the validation tests, a total of 13 algorithms and their respective implementations were selected. Specifically, the algorithms used were:

\begin{figure*}[!hb]
    \centering
    \includegraphics[width=1\textwidth]{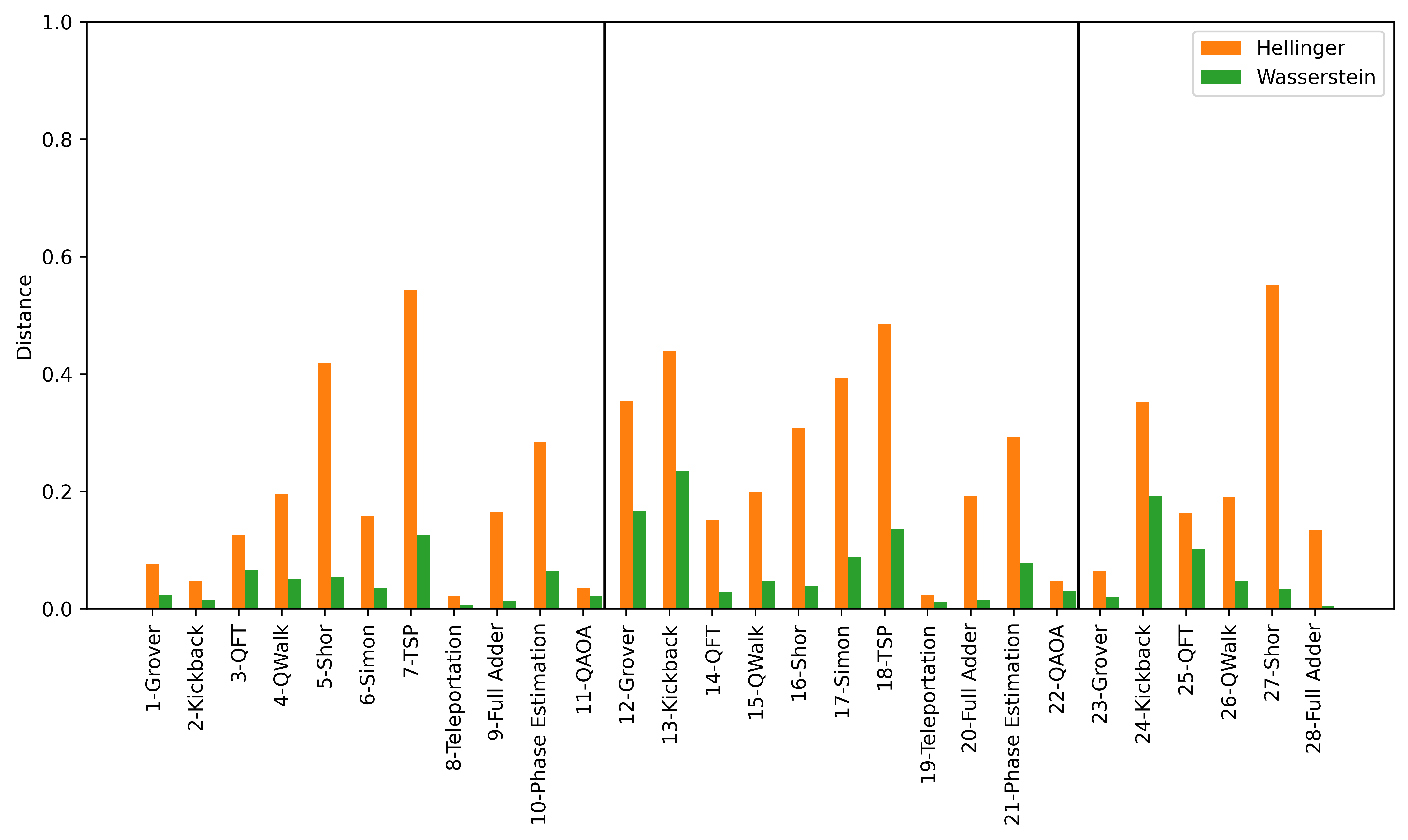}
    \caption{Distance between schedule and individual execution}
    \label{fig:hellingerwasserstein}
\end{figure*}

\begin{itemize}
\item Berstein-Vazirani;
\item Deutsch-Jozsa;
\item Full Adder;
\item Grover;
\item Kickback;
\item Phase Estimation;
\item Quantum Approximate Optimization Algorithm (QAOA);
\item Quantum Fourier Transform (QFT);
\item Quantum Walk;
\item Shor;
\item Simon;
\item Teleportation;
\item Travelling Salesman Problem (TSP).
\end{itemize}

These algorithms cover a wide range of applications and relevant problems in quantum computing, allowing us also to evaluate the effectiveness of the proposal in different contexts.

It should be noted that some of these algorithms, due to their non-deterministic nature, could not be taken into account when performing the final validation as explained in Section \ref{Threats} of threats to validity.

%To validate the proposal, a total of 3 schedules have been made in order to reach the maximum capacity of 127 qubits offered by the quantum processor. The first two schedules contain the thirteen complete circuits, while the last one contains eight, which are the ones that could be added with the remaining qubits of the first two schedules. 

To validate the proposal, a set of 13 circuits has been sent to execute, 3 times each circuit for a total of 39 execution tasks. Given the 127 qubits offered by the quantum processor, the scheduler generated a combined circuit including 34 of the 39 tasks. Two full sets of the 13 circuits plus an additional 8 circuits for the remaining qubits. For the execution of this 34 combined circuits and the individual execution of the 13 original circuits, 10.000 shots were used on the IBM ``Eagle'' processor.

\subsection{Noise Results}

Figure \ref{fig:quantum_algorithms} shows some of the most representative obtained results. These graphs show the obtained results in scheduled executions versus the individual results. In many of the obtained graphs, these results do not differ greatly and it is possible to determine the correct results given their amplitude. 

The results of the algorithm in its execution within the scheduling are shown in blue, the results of the individual execution in yellow and the coincidence between both results in brown. In most of the obtained results, the individual runs are very similar to the scheduled runs, as can be seen in Figure \ref{fig:qaoa1}, \ref{fig:simon1} and \ref{fig:grover1}. There are some cases such as Shor's algorithm---Figure \ref{fig:shor1}---where the results in scheduling are not quite close to the expected results. In the latter case, this is due to the circuit design of Shor's algorithm, which consists of a large number of CNOT gates. However, even though the results of the scheduling have suffered from a different amount of noise, it is still possible to distinguish which is the correct value because it is still the one that shows a higher frequency, such as the results shown in Figure \ref{fig:fullAdder}. 

%Para validar formalmente los resultados obtenidos se han realizado tres estudios estadísticos entre los resultados obtenidos por cada cirucito ejecutado individualmente y en composición. %frente a los resultados esperados de la ejecución sin ruido. Para ello, se han utilizado tres algoritmos estadísticos para medir la distancia con distribuciones de probabilidad, Total Variation Distance, Hellinger Distance y Wasserstein Distance \cite{cai2022distances}. Estos tres algoritmos permiten obtener un valor entre 0 y 1 que indica la diferencia entre dos distribuciones, mientras menor sea el valor, más similares son. Para comprobar qué opción es mejor entre los resultados individuales y en composición, se ha obtenido la diferencia entre las distancias obtenidas en ambos casos, para observar si difieren en gran cantidad.

In order to formally validate the results obtained, two statistical studies were performed between the results obtained by each circuit executed individually and in scheduling. For this purpose, two statistical algorithms have been used to measure the distance with probability distributions: Hellinger Distance, which measures the statistical difference between the results, and Wasserstein Distance, which measures how much it costs to convert one distribution into the other \cite{cai2022distances}. These two algorithms allow to obtain a value between 0 and 1 that indicates the difference between two distributions, the smaller the value, the more similar they are. To check which option is better and to see if they differ by a large amount, the difference between the distances has been obtained.

%La Figura \ref{fig:hellingerwasserstein} muestra la distancia entre las ejecuciones en composición frente a las ejecuciones individuales de los algoritmos seleccionados. Las barras naranjas corresponden a la distancia utilizando el algoritmo de Hellinger mientras que las barras verdes indican la distancia según el algoritmo de Wasserstein. Se observa que la mayor parte de los circuitos en composición tienen una distancia bastante baja frente a las ejecuciones individuales, esto significa que el ruido y la decoherencia ha afectado de forma leve a la composición. Siendo el porcentaje de diferencia un 23\% con el algoritmo de Hellinger y únicamente un 6\% con el algoritmo de Wasserstein.

Figure \ref{fig:hellingerwasserstein} shows the distance between the schedule executions versus the individual executions of the selected algorithms. The orange bars correspond to the distance using the Hellinger algorithm while the green bars indicate the distance according to the Wasserstein algorithm. It is observed that most of the circuits in scheduling have a rather low distance versus the individual executions, this means that noise and decoherence has slightly affected the scheduling. Overall, the percentage difference is 23\% with Hellinger's algorithm and only 6\% with Wasserstein's algorithm, although there are cases such as \textit{8-Teleportation} or \textit{28-Full Adder} where the percentage with Wasserstein is practically 0\%. 

If we take as an example the case of \textit{27-Shor}, the Hellinger distance obtained between the scheduling and the individual performance is 0.6, being this the case that differs the most, although as discussed with Figure \ref{fig:shor1}, it still gives the expected result. Most cases, such as \textit{1-Grover} or \textit{8-Teleportatio}n, have a minimal distance between both distributions of results, indicating that they have suffered less noise. 

%En la mayoría de circuitos las ejecuciones individuales han obtenido mejores resultados, pero en una gran parte de los circuitos, esta mejora no es tan significativa. A su vez, existen algunos circuitos que han obtenido mejores resultados al realizarse la ejecución dentro de la composición y no se ven afectados gravemente por la decoherencia ni el ruido. 

%\textcolor{red}{- reducción de costes, tiempo de cola y límite de ejeución de tres tareas paralelas en IBM} 

After validation, it can be concluded that the scheduling of the circuits allows an improvement in the queuing times. Moreover, it has been possible to execute 34 circuits by taking advantage of the 100\% of the machine.

Another advantage of using the scheduler is the reduction of execution costs. The sum of the time of the individual executions was 70 seconds, while the scheduling took 24 seconds. The execution time of the circuit in scheduling is very close to the time it takes to execute the deepest circuit. In this way, it has been possible to greatly reduce this time, which allows a significant cost reduction.

On the other hand, IBM only allows the execution of 3 circuits per user at the same time, with the use of scheduling it is possible to increase the number of circuits that can be executed at the same time.

%%%%%%%%%%%%%%%%%%%%%%%%%%%%%%%%%%%%%%%%%%%%%%%%%%%%%%%%%%%%%%%%%%%%%%%%%%%%%%

\subsection{Threats to validity}\label{Threats}

Despite the results obtained in this study, it is important to take into account certain threats to validity.

First of all, some non-deterministic algorithms have not been included in the final validation process. For example, algorithms such as Bernstein-Vazirani or Deutsch-Jozsa do not allow us to perform the validation correctly as they cannot establish what the expected results would be and, therefore, they introduce problems when comparing individual and schedule executions. 

In addition, variations in execution times and calibration of quantum computers may introduce uncertainties that could affect the conclusions of the study. It is important to recognize the possible influence of these factors on the results and their interpretation. Also, it should be noted that the variability of quantum environments introduces a level of uncertainty that can difficult make interpretation of the data, hence the use of statistical analysis to validate the results. 

Finally, the study was performed on the quantum processors offered by IBM, as these are the ones that currently offer the highest number of qubits. However, it should be noted that this aspect may limit the generalization of the results to other cloud service providers, with different configurations and topologies.

%%%%%%%%%%%%%%%%%%%%%%%%%%%%%%%%%%%%%%%%%%%%%%%%%%%%%%%%%%%%%%%%%%%%%%%%%%%%%%%%%%%%%%%%%
\section{Conclusion And Future Work}\label{sec:conclusion}

%\textcolor{red}{- conclusión del trabajo}

This work has addressed some of the current constraints observed in quantum computing, such as the high demand for quantum computational resources, resulting in long waiting times, restrictions on qubit availability and high costs for developers using these technologies. In addition, inherent limitations of quantum computers have been identified, such as the presence of interference between qubits, calibration problems and various noise sources.

To address these problems, an approach has been proposed to reduce waiting times and optimize task execution in quantum computers by scheduling circuits from different users. This proposal is based on the idea of unifying circuits in an efficient way to increase the efficiency in the utilization of the available resources in quantum computers, allowing to make a better use of the qubits offered by these machines.

The main challenge of the study has been to manage the errors that may occur in the results, but as it has been proved, the distribution of the results in scheduling versus individual execution is very similar. The results obtained indicate that the noise generated by using a number of qubits close to the maximum number allowed by the machine does not affect the results. 

%\textcolor{red}{- resumir las mejoras obtenidas, porcentaje de mejora...}

In summary, executing circuits in scheduling has allowed an improvement in waiting times, with a considerable reduction in execution time compared to individual executions. For example, the total execution time was reduced from 70 seconds in individual runs to 24 seconds in scheduling. Thus, by reducing the execution time, this translates into a reduction in operating costs, not including queuing times.

Furthermore, despite the presence of noise and decoherence, the results obtained in scheduling have shown a remarkable similarity with the ideal results. This suggests that the noise generated by using a number close to the maximum number of qubits does not significantly affect the results.

%Por ello, todos los resultados obtenidos permiten continuar con esta línea de investigación tras conocer que los beneficios para los desarrolladores de software cuántico son visibles.

%Therefore, all the results obtained allow us to continue with this line of research after knowing that the benefits for quantum software developers are visible.

%\textcolor{red}{- trabajos futuros (más poroveedores, coordinación del número de shots, combinación api con quirk, composición de resultados..., green computing, poder editar el circuito cuando está en cola)}

Thus, several areas of research for future work are identified that could expand and improve the proposal. One of these areas is the design of a circuit scheduling optimization algorithm that addresses different policies and criteria. For this proposal, it has been considered to assign the same number of shots per circuit. Therefore, we are exploring the possibility of extending this option as one of the policies for scheduling, so that each user can specify the number of shots desired, independently of the choices of other developers.

%It is also proposed to implement a functionality that allows users to edit circuits while they are queued for execution. This would provide greater flexibility and control over the circuit composition and execution processes.

Finally, to evaluate the effectiveness of the proposal in different environments, it is proposed to perform additional testing and validation using a variety of cloud providers of quantum services.

\section*{Acknowledgments}
%Anonymous acknowledgements

This work has been partially funded by the European Union “Next GenerationEU /PRTR”, by the Ministry of Science, Innovation and Universities (projects PID2021-1240454OB-C31, TED2021-130913B-I00, and PDC2022-133465-I00). It is also supported by QSERV: Quantum Service Engineering: Development Quality, Testing and Security of Quantum Microservices project funded by the Spanish Ministry of Science and Innovation and ERDF; by the Regional Ministry of Economy, Science and Digital Agenda of the Regional Government of Extremadura (GR21133); and by European Union under the Agreement - 101083667 of the Project “TECH4E -Tech4effiency EDlH” regarding the Call: DIGITAL-2021-EDlH-01 supported by the European Commission through the Digital Europe Program. It is also supported by grant PRE2022-102070, funded by  MCIN/AEI/10.13039/501100011033 and by FSE+.

%%%%%%%%%%%%%%%%%%%%%%%%%%%%%%%%%%%%%%%%%%%%%%%%%%%%%%%%%%%%%%%%%%%%%%%%%%%%%%%%%%%%%%%%
\bibliographystyle{IEEEtranDOI}
\bibliography{references}

\begin{thebibliography}{10}
\providecommand{\url}[1]{#1}
\csname url@samestyle\endcsname
\providecommand{\newblock}{\relax}
\providecommand{\bibinfo}[2]{#2}
\providecommand{\BIBentrySTDinterwordspacing}{\spaceskip=0pt\relax}
\providecommand{\BIBentryALTinterwordstretchfactor}{4}
\providecommand{\BIBentryALTinterwordspacing}{\spaceskip=\fontdimen2\font plus
\BIBentryALTinterwordstretchfactor\fontdimen3\font minus \fontdimen4\font\relax}
\providecommand{\BIBforeignlanguage}[2]{{%
\expandafter\ifx\csname l@#1\endcsname\relax
\typeout{** WARNING: IEEEtran.bst: No hyphenation pattern has been}%
\typeout{** loaded for the language `#1'. Using the pattern for}%
\typeout{** the default language instead.}%
\else
\language=\csname l@#1\endcsname
\fi
#2}}
\providecommand{\BIBdecl}{\relax}
\BIBdecl

\bibitem{hassija2020present}
V.~Hassija, V.~Chamola, V.~Saxena, V.~Chanana, P.~Parashari, S.~Mumtaz, and M.~Guizani, ``\href{https://doi.org/10.1049/iet-qtc.2020.0027}{{Present landscape of quantum computing}},'' \emph{IET Quantum Communication}, vol.~1, no.~2, pp. pp. 42--48, 2020.

\bibitem{islam2015review}
M.~M. Islam and M.~Rahaman, ``{A review on progress and problems of quantum computing as a service (QCaaS) in the perspective of cloud computing},'' \emph{Global Journal of Computer Science and Technology}, vol.~15, no.~B4, pp. 23--26, 2015.

\bibitem{alvarado2023technological}
J.~Alvarado-Valiente, J.~Romero-{\'A}lvarez, E.~Moguel, J.~Garc{\'\i}a-Alonso, and J.~M. Murillo, ``\href{https://doi.org/10.1007/s11219-023-09633-5}{Technological diversity of quantum computing providers: a comparative study and a proposal for api gateway integration},'' \emph{Software Quality Journal}, pp. 1--21, 2023.

\bibitem{choi2023ibm}
C.~Q. Choi, ``\href{https://doi.org/10.1109/MSPEC.2023.10006669}{Ibm's quantum leap: The company will take quantum tech past the 1,000-qubit mark in 2023},'' \emph{IEEE Spectrum}, vol.~60, no.~1, pp. 46--47, 2023.

\bibitem{ichikawa2023comprehensive}
T.~Ichikawa, H.~Hakoshima, K.~Inui, K.~Ito, R.~Matsuda, K.~Mitarai \emph{et~al.}, ``\href{https://doi.org/10.48550/arXiv.2307.16130}{A comprehensive survey on quantum computer usage: How many qubits are employed for what purposes?}'' \emph{arXiv:2307.16130}, 2023.

\bibitem{nguyen2024qfaas}
H.~T. Nguyen, M.~Usman, and R.~Buyya, ``\href{https://doi.org/10.1016/j.future.2024.01.018}{Qfaas: A serverless function-as-a-service framework for quantum computing},'' \emph{Future Generation Computer Systems}, vol. 154, pp. 281--300, 2024.

\bibitem{garg2002techniques}
R.~P. Garg, I.~A. Sharapov, and I.~Sharapov, \emph{Techniques for optimizing applications: high performance computing}.\hskip 1em plus 0.5em minus 0.4em\relax Sun Microsystems Press Palo Alto, 2002.

\bibitem{moguel2022quantum}
E.~Moguel, J.~Rojo, D.~Valencia, J.~Berrocal, J.~Garcia-Alonso, and J.~M. Murillo, ``\href{https://doi.org/10.1007/s11219-022-09589-y}{Quantum service-oriented computing: current landscape and challenges},'' \emph{Software Quality Journal}, vol.~30, no.~4, pp. 983--1002, 2022.

\bibitem{zhao2020quantum}
J.~Zhao, ``\href{https://doi.org/10.48550/arXiv.2007.07047}{Quantum software engineering: Landscapes and horizons},'' \emph{arXiv preprint arXiv:2007.07047}, 2020.

\bibitem{mccaskey2018hybrid}
A.~McCaskey, E.~Dumitrescu, D.~Liakh, and T.~Humble, ``\href{https://doi.org/10.1109/ICRC.2018.8638598}{Hybrid programming for near-term quantum computing systems},'' in \emph{2018 IEEE international conference on rebooting computing (ICRC)}.\hskip 1em plus 0.5em minus 0.4em\relax IEEE, 2018, pp. 1--12.

\bibitem{ravi2021quantum}
G.~S. Ravi, K.~N. Smith, P.~Gokhale, and F.~T. Chong, ``\href{https://doi.org/10.48550/arXiv.2203.13121}{Quantum computing in the cloud: Analyzing job and machine characteristics},'' in \emph{2021 IEEE International Symposium on Workload Characterization (IISWC)}.\hskip 1em plus 0.5em minus 0.4em\relax IEEE, 2021, pp. 39--50.

\bibitem{garcia2021quantum}
J.~Garcia-Alonso, J.~Rojo, D.~Valencia, E.~Moguel, J.~Berrocal, and J.~M. Murillo, ``\href{https://doi.org/10.1109/MIC.2021.3132688}{Quantum software as a service through a quantum api gateway},'' \emph{IEEE Internet Computing}, vol.~26, no.~1, pp. 34--41, 2021.

\bibitem{Nguyen2023obm}
H.~T. Nguyen, M.~Usman, and R.~Buyya, ``\href{https://doi.org/10.1109/QSW59989.2023.00013}{{iQuantum: A Case for Modeling and Simulation of Quantum Computing Environments}},'' in \emph{{2023 IEEE International Conference on Quantum Software}}.\hskip 1em plus 0.5em minus 0.4em\relax Los Alamitos, CA, USA: IEEE Computer Society, 3 2023, pp. 21--30.

\bibitem{leymann2020bitter}
F.~Leymann and J.~Barzen, ``\href{https://doi.org/10.1088/2058-9565/abae7d}{The bitter truth about gate-based quantum algorithms in the nisq era},'' \emph{Quantum Science and Technology}, vol.~5, no.~4, p. 044007, 2020.

\bibitem{brandt1999qubit}
H.~E. Brandt, ``\href{https://doi.org/10.1016/S0079-6727(99)00003-8}{Qubit devices and the issue of quantum decoherence},'' \emph{Progress in Quantum Electronics}, vol.~22, pp. 257--370, 1999.

\bibitem{murali2020software}
P.~Murali, D.~C. McKay, M.~Martonosi, and A.~Javadi-Abhari, ``\href{https://doi.org/10.1145/3373376.3378477}{Software mitigation of crosstalk on noisy intermediate-scale quantum computers},'' in \emph{International Conference on Architectural Support for Programming Languages and Operating Systems}, 2020, pp. 1001--1016.

\bibitem{nation2021scalable}
P.~D. Nation, H.~Kang, N.~Sundaresan, and J.~M. Gambetta, ``\href{https://doi.org/10.1103/PRXQuantum.2.040326}{Scalable mitigation of measurement errors on quantum computers},'' \emph{PRX Quantum}, vol.~2, no.~4, p. 040326, 2021.

\bibitem{wilson2020just}
E.~Wilson, S.~Singh, and F.~Mueller, ``\href{https://doi.org/10.1109/QCE49297.2020.00050}{Just-in-time quantum circuit transpilation reduces noise},'' in \emph{2020 IEEE international conference on quantum computing and engineering (QCE)}.\hskip 1em plus 0.5em minus 0.4em\relax IEEE, 2020, pp. 345--355.

\bibitem{johnstun2021understanding}
S.~Johnstun and J.-F. Van~Huele, ``\href{https://doi.org/10.1119/10.0006204}{Understanding and compensating for noise on ibm quantum computers},'' \emph{American Journal of Physics}, vol.~89, no.~10, pp. 935--942, 2021.

\bibitem{usandizaga2023quantum}
E.~M. Usandizaga, T.~Yue, P.~Arcaini, and S.~Ali, ``\href{https://doi.org/10.48550/arXiv.2311.16913}{Which quantum circuit mutants shall be used? an empirical evaluation of quantum circuit mutations},'' \emph{arXiv preprint arXiv:2311.16913}, 2023.

\bibitem{kaewpuang2023stochastic}
R.~Kaewpuang, M.~Xu, D.~Niyato, H.~Yu, Z.~Xiong, and J.~Kang, ``\href{https://doi.org/10.1109/NOMS56928.2023.10154430}{Stochastic qubit resource allocation for quantum cloud computing},'' in \emph{NOMS 2023-2023 IEEE/IFIP Network Operations and Management Symposium}.\hskip 1em plus 0.5em minus 0.4em\relax IEEE, 2023, pp. 1--5.

\bibitem{chirolli2008decoherence}
L.~Chirolli and G.~Burkard, ``\href{https://doi.org/10.1080/00018730802218067}{Decoherence in solid-state qubits},'' \emph{Advances in Physics}, vol.~57, no.~3, pp. 225--285, 2008.

\bibitem{cai2022distances}
Y.~Cai and L.-H. Lim, ``\href{https://doi.org/10.1109/TIT.2022.3148923}{Distances between probability distributions of different dimensions},'' \emph{IEEE Transactions on Information Theory}, vol.~68, no.~6, pp. 4020--4031, 2022.

\end{thebibliography}

\end{document}